\documentclass[aps,prb,preprint]{revtex4}

\usepackage{graphicx}

\usepackage{keyval}

\usepackage{amsmath}

\setkeys{Gin}{width=0.750\columnwidth}

\begin{document}

\preprint{ERAU-PHY-0201c}

\title{The driven pendulum at arbitrary drive angles}

\author{Gordon J.\ VanDalen}

\email{gordonvandalen@aol.com}

\affiliation{Department of Physics, University of California,
Riverside, California 92521}

\affiliation{Department of Physics, Embry-Riddle Aeronautical
University, Prescott, Arizona 86301\\
Received: 11/8/02  Accepted: 6/20/03  Index Code: 3.20}

\date{\today}

\begin{abstract}
We discuss the equation of motion of the driven pendulum and
generalize it to arbitrary driving angle. The pendulum will
oscillate about a stable angle other than straight down if the
drive amplitude and frequency are large enough for a given drive
angle. The emphasis is on the parameters associated with a simply
made demonstration apparatus.
\end{abstract}

\maketitle

\section{\label{sec:intro} Introduction}

The general theory of the driven inverted pendulum with the drive angle
restricted to 180$^\circ$ is given in Ref.~\onlinecite{Landau} and has
been discussed in Refs.~\onlinecite{FMP,DCM,EIB}; the latter reference
includes a good introduction to the physical system and further
citations. In the following, the behavior of the harmonically driven
pendulum will be described for any drive angle from the vertical.

An inverted pendulum demonstration\cite{Jones} that is designed to be
clamped to a table top becomes more interesting when hand held. We will
examine the stability of this system as a function of the angular
frequency, the drive amplitude, and the drive angle from the vertical. 
This type of demonstration is best used in a junior level
classical mechanics course when introducing Lagrange's equations to find
the equations of motion. Examples of driving the support of a simple
pendulum harmonically can be introduced, but only to set up the
equation of motion.\cite{Marion,Symon} The same apparatus can be used in
an advanced graduate classical mechanics course where the harmonically
driven pendulum is used as an example of Lagrange's equations and as a
source of problems.\cite{Landau}

While holding the saw in my hand,\cite{m17q} with the power still
on, I lowered the saw and observed the peculiar behavior of the
pendulum as it sometimes found new stable angles of oscillation as
the drive angle changed. Changing the driving angle in the plane
of the pendulum's motion introduces the new and interesting
problem that I address in this paper.

\section{Equation of Motion of the Pendulum Driven at Any Angle}
\begin{figure}[h]
\includegraphics[width=0.5\columnwidth]{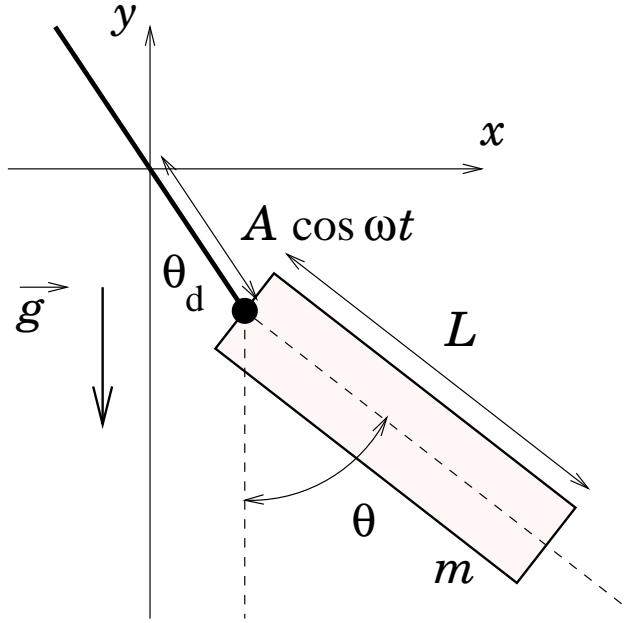} 
\caption{\label{fig:geom}The pendulum driven at angle $\theta_d$.}
\end{figure}

The dynamics of a driven inverted pendulum near $\theta =
180^\circ$ are described in
Refs.~\onlinecite{Landau,FMP,DCM,EIB}. We now generalize the
problem to arbitrary driving angles. The geometry of the problem is
illustrated in Fig.~\ref{fig:geom}. A thin uniform rod of length
$L$ is driven at one end (pivot) with amplitude
$A$, angular frequency $\omega$, at an angle of $\theta_d$ from the
downward vertical.
The angle of the
thin rod is measured by the generalized coordinate $\theta$ which
also is measured from the vertical downward. (We also could have
used a simple pendulum consisting of a mass $m$ at the end of a
light rod of length
$\ell$, or a general pendulum with moment of inertia $I$ and center
of mass at a distance
$z$ from the pivot. The rod pendulum geometry chosen here
corresponds to a simple lecture apparatus.\cite{Jones,m17q} A
different pendulum geometry can be easily substituted into the
equations of motion below.)

The angle of the rod $\theta$ could be measured from the driving
direction,
$\theta_d$. However, for only two values of $\theta_d$ (0 and
$\pi$) will the equilibrium angle of the pendulum be $\theta_d$.
For any value of
$\theta_d$, there is a stable point for small $\omega$ and $A$ at
$\theta = 0$. So we choose to measure the pendulum angle $\theta$
from the downward vertical.

The kinetic energy and potential energy of the rod can be written
in terms of the generalized coordinate $\theta$. The energies are
separated into the motion of the center of mass plus the rotation
about the center of mass. We first express the Cartesian
coordinates and velocities of the center
of mass in terms of $\theta$.
\begin{subequations}
\label{eqn:xv}
\begin{eqnarray}
x_{\text{cm}}&=& 
+{L\over 2}\sin\theta +A\sin\theta_d \cos\omega t \\
y_{\text{cm}} &=& 
-{L\over 2} \cos\theta - A \cos\theta_d \cos\omega t
\label{eqn:xvdot} \\
\dot{x}_{\text{cm}} &=& 
+{L\over 2}\dot{\theta} \cos\theta -A
\omega\sin\theta_d\sin\omega t\\
\dot{y}_{\text{cm}} &=& 
+{L\over 2}\dot{\theta} \sin\theta +A 
\omega\cos\theta_d \sin\omega t.
\end{eqnarray}
\end{subequations}

We can express the kinetic energy and potential energy in terms of the
generalized coordinate $\theta$.
\begin{equation}
T = {1\over 2} m ( \dot{x}_{\text{cm}}^2+\dot{y}_{\text{cm}}^2 ) +
{1\over 2} I_{\text{cm}}\dot{\theta}^2 .
\end{equation}
If we use the Cartesian coordinates and velocities from
Eq.~(\ref{eqn:xv}), we obtain
\begin{equation}
T = {1\over 2} \big [ {{m L^2}\over{3}} \dot{\theta}^2 +
m L A\omega \dot{\theta} (\sin\theta \cos\theta_d - 
\cos\theta \sin\theta_d ) \sin\omega t + 
m A^2 \omega^2 \sin^2 \omega t \big] . \label{eqn:T}
\end{equation}
The first term on the right-side of Eq.~(\ref{eqn:T}) 
corresponds to the rotation about the moving support at the end of
the pendulum with a moment of inertia of $m L^2/3$ about that end.

The potential energy depends only on $y_{\text{cm}}$, and from
Eq.~(\ref{eqn:xv}) we have
\begin{equation}
V = m g y_{\text{cm}} = 
- {{m g L}\over 2} \cos\theta - m g A \cos\theta_d \cos\omega t .
\label{eqn:V}
\end{equation}
The Lagrangian is ${\cal L} = T-V$, and using
Eqs.~(\ref{eqn:T}) and (\ref{eqn:V}), we have
\begin{eqnarray}
{\cal L} &=& {1\over 2} \big[ {{m L^2}\over{3}} \dot{\theta}^2 +
m L A\omega \dot{\theta} \sin\big( \theta - \theta_d\big) 
\sin\omega t 
+ m A^2 \omega^2 \sin^2 \omega t \big] \nonumber \\
&&{} + mg \big ({L\over 2}\cos\theta +A\cos\theta_d\cos\omega t\big) .
\label{eqn:L}
\end{eqnarray}
The equation of motion is found from Lagrange's equation for the
single generalized coordinate $\theta$:
\begin{equation}
{d \over{dt}}\big( {{\partial {\cal L}}\over{\partial
\dot\theta}}\big) - {{\partial {\cal L}}\over{\partial \theta}} = 0 .
\end{equation}
If we evaluate the derivatives of the Lagrangian from
Eq.~(\ref{eqn:L}), we obtain the equation of motion for a pendulum
driven at any angle:
\begin{equation}
\ddot{\theta} + {{3 A\omega^2}\over{2 L}} 
\sin ( \theta - \theta_d) \cos\omega t +
{{3 g}\over{2 L}} \sin\theta = 0.
\label{eqn:motion}
\end{equation} Note that if $\theta_d = 180^\circ$, we obtain the
usual driven inverted pendulum.\cite{Landau,FMP,DCM,EIB} This
special case is revisited briefly in Sec.~\ref{sec:INV} and in the
Appendix.

\section{\label{sec:effV}Effective potential}
We now introduce an effective potential to help us understand the
physical origin of the stability of the inverted pendulum and to
parameterize the condition for stability at any driving angle.

Landau and Lifshitz\cite{Landau} separated the motion of the
horizontally or vertically driven pendulum into two parts: a
\textit{fast} component
$\xi(t)$ at the drive angular frequency $\omega$, and a
\textit{slow} component $\phi(t)$ that describes the slower
overall swinging of the driven pendulum:
\begin{equation}
\theta(t) = \phi(t) + \xi(t)
\end{equation}
The angle $\phi$ is defined as zero downward as is $\theta$, and
$\xi$ is the difference between $\theta$ and $\phi$.
In what follows we assume that the average 
of $\xi(t)$ (denoted as $\overline{\xi}$) is zero, and that
$\xi(t)$ is small.

The equation of motion for a pendulum driven 
at any angle, Eq.~(\ref{eqn:motion}), can be rearranged, in the form
\begin{equation}
\ddot{\theta} = - {{3 g}\over{2 L}} \sin\theta - 
{{3 A\omega^2}\over{2 L}} 
\sin (\theta-\theta_d ) \cos\omega t = 
F(\theta) + f(\theta, t) .
\label{eqn:sep}
\end{equation} We have separated the angular acceleration into a
time dependent driving term $f(\theta,t)$ oscillating at the
driving angular frequency $\omega$, and a time independent part
$F(\theta)$ corresponding to the
effect of gravity. Equation~(\ref{eqn:sep}) can be rewritten in an
expansion in
$\xi$ for small values of $\xi(t)$:
\begin{equation}
\label{keep}
\ddot{\phi} + \ddot{\xi} \simeq 
F(\phi) + {{dF}\over{d\theta}}(\phi ) \xi
+ f(\phi,t) + {{df}\over{d\theta}}(\phi,t) \xi.
\end{equation}
We keep only the largest rapidly varying terms on each side of
Eq.~(\ref{keep}) and write
\begin{equation}
\ddot{\xi} \simeq f(\phi,t) = - {{3 A\omega^2}\over{2 L}} 
\sin(\phi-\theta_d) \cos\omega t .
\label{eqn:ximotion}
\end{equation}
Note that $\ddot{\xi}$ is $\omega^2$ larger than $\xi(t)$, so
the terms in $\xi$ may safely be ignored. Also both $\ddot{\phi}$
on the left and $F(\theta)$ on the right do not oscillate at the
driving angular frequency $\omega$.

Equation~(\ref{eqn:ximotion}) may be integrated twice, taking
the slow motion
$\phi$ to be constant on the time scale $1/\omega$:
\begin{equation}
\xi(t) \simeq {{3 A}\over{2 L}} 
\sin(\phi-\theta_d) \cos\omega t .
\label{eqn:xi}\end{equation}
By averaging Eq.~(\ref{eqn:sep}) over the 
fast component of the motion at angular frequency
$\omega$, we obtain the equation of motion for the slow swinging of
the pendulum:
\begin{equation}
\ddot{\phi} + \overline{\ddot{\xi}} \simeq 
F(\phi) + {{dF(\phi)}\over{d\theta}} \overline{\xi}
+ \overline{f(\phi,t)} + 
\overline{{{df(\phi,t)}\over{d\theta}} \xi} .
\end{equation}
The rapidly oscillating terms $\ddot{\xi}$, $\xi$, and $f(\phi,t)$
average to zero, and we have
\begin{equation}
\ddot{\phi} \simeq F(\phi) + 
\overline{{{df}\over{d\theta}}(\phi,t) \xi} .
\end{equation}
We then substitute for $F(\phi)$ and $df/dt$ from 
Eq.~(\ref{eqn:sep}) and for $\xi(t)$ from Eq.~(\ref{eqn:xi}).
Only the $\cos\omega t$ terms vary rapidly, and so can be averaged on the
longer time scale of the change of $\phi$. If
we use $\overline{\cos^2\omega t} = {1\over 2}$, we obtain the
equation of motion for the angle $\phi$:
\begin{equation}
\ddot{\phi} \simeq - {{3 g}\over{2 L}} \sin\phi - 
{{9 A^2 \omega^2}\over{8 L^2}} \cos(\phi-\theta_d)
\sin(\phi-\theta_d).
\label{eqn:ddotphi}
\end{equation}
Equation~(\ref{eqn:ddotphi}) describes the slow swinging motion of
the driven pendulum. If we use a simple trigonometric identity, we
obtain another useful form of
the equation of motion:
\begin{equation}
\ddot{\phi} \simeq - {{3 g}\over{2 L}} \sin\phi - 
{{9 A^2 \omega^2}\over{16 L^2}}
\sin [ 2 (\phi-\theta_d) ] .
\label{eqn:ddotphi2}
\end{equation}

The effective torque is the acceleration $\ddot{\phi}$ about one
end of the rod multiplied by the moment of inertia about the end
of the rod:
\begin{equation}
\tau_{\text{eff}}(\phi) = {{m L^2}\over 3} \ddot{\phi}
\simeq - {{m g L}\over{2}} \big(\sin\phi +
{{3 A^2 \omega^2}\over{8 g L}} 
\sin[2 (\phi-\theta_d)]\big ).
\label{eqn:torque}
\end{equation}
This torque can be derived from an effective potential 
energy, $V_{\text{eff}}(\phi)$, where 
$\tau_{\text{eff}}(\phi) = - dV_{\text{eff}}/d\phi$.
\begin{equation} 
V_{\text{eff}}(\phi) = 
- {{m g L}\over{2}} \big(\cos\phi +
{{3 A^2 \omega^2}\over{16 g L}}
\cos [2(\phi-\theta_d)]\big) .
\label{eqn:Veff}
\end{equation}
We define the dimensionless critical parameter, $R$, and the critical
angular frequency, $\omega_c$:
\begin{subequations}
\label{eqn:Rdef}
\begin{eqnarray}
R &=& {{3 A^2 \omega^2}\over{4 g L}} =
{{\omega^2}\over{\omega_c^2}}\\
\omega_c &=& 
\sqrt{ {{4 g L}\over{3A^2}}} ,
\end{eqnarray}
\end{subequations}
and rewrite the effective potential from Eq.~(\ref{eqn:Veff}):
\begin{equation}
V_{\text{eff}}(\phi) = 
- {{m g L}\over{2}} \big(\cos\phi + {R\over 4}
\cos[2(\phi-\theta_d)]\big) .
\label{eqn:VeffR}
\end{equation}
The first term on the right side of Eq.~(\ref{eqn:VeffR}) is
simply the effect of gravity acting on the center of mass of the
pendulum. The second term comes from the dynamics of the forced
motion, and represents the average kinetic energy of the rapidly
driven oscillation of the pendulum about its center of mass. As the
pendulum deviates from the drive angle $\theta_d$, the angular
kinetic energy of the pendulum about its center of mass increases.

\begin{figure}[h]
\includegraphics{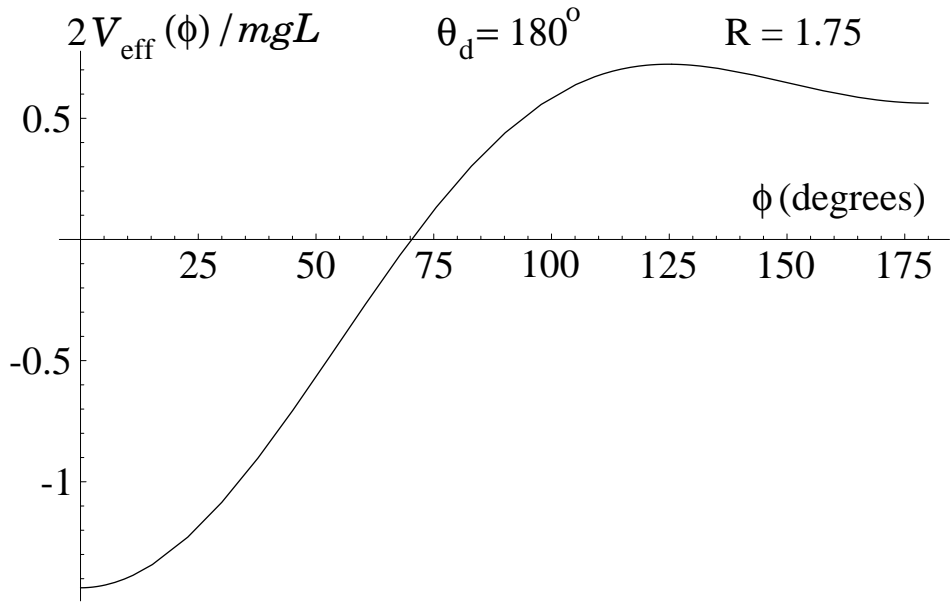}
\caption{\label{fig:Vinv} The potential $V_{\text{eff}}(\phi)$ for
a driven inverted pendulum ($\theta_d = 180^\circ$).}
\end{figure}

If treated as an effective potential energy, the kinetic energy
associated with the driving angular frequency stabilizes the
slow motion of the inverted pendulum. Figure~\ref{fig:Vinv} shows
the effective potential as a function of $\phi$ for a driven
inverted pendulum with $R=1.75$, corresponding to the apparatus of
Ref.~\onlinecite{m17q}. In Fig.~\ref{fig:Vinv} we see the
gravitational potential minimum at $\phi=0$ and also the
dynamic potential minimum at $\phi = 180^\circ$. If the
drive amplitude or angular frequency become too small ($R \le
1$), then the stable equilibrium at $\phi = 180^\circ$
disappears. (Note that the local maximum near 125 degrees limits
the amplitude of slow oscillation of this particular case of a
driven inverted pendulum.)

The same physical interpretation of the driven wobble of the
pendulum as a stabilizing effective torque is seen in
Eq.~(\ref{eqn:torque}), which includes a stabilizing term
proportional to $\omega^2$ whose origin is the
kinetic energy of rotation at the driving angular frequency
$\omega$. (This term is analogous to the
centrifugal force in orbital motion which originates from the
rotational kinetic energy about the center of mass when expressed
as a radial equation of motion.) Reference~\onlinecite{EIB} offers
additional physical insight into the stability of the inverted
pendulum.

\section{Special Cases of the Drive Angle}
Let us look at three special cases before moving on to general values
of $\theta_d$. These three examples will provide guidance in 
interpreting the result for arbitrary values of $\theta_d$.

\subsection{Drive angle zero degrees}
If $\theta_d = 0$, we obtain the equation of motion for the
slow oscillation $\phi(t)$ from Eq.~(\ref{eqn:ddotphi}):
\begin{equation}
\ddot{\phi} + \big({{3 g}\over{2 L}} + 
{{9 A^2 \omega^2}\over{8 L^2}} \cos\phi \big) \sin\phi = 0 .
\end{equation}
For small angles $\phi \sim 0$, we can take $\cos\phi\simeq1$ and 
$\sin\phi \simeq \phi$:
\begin{equation}
\ddot{\phi} + {{3 g}\over{2 L}}\big( 1 + 
{{3 A^2 \omega^2}\over{4 g L}}\big) \phi = 
\ddot{\phi} + \omega_p^2 \phi \simeq 0 .
\end{equation}
The pendulum oscillates slowly about $\phi = 0$ with an angular
frequency $\omega_p$, which is the square root of the coefficient
of the term in $\phi(t)$:
\begin{equation}
\omega_p = 
\omega_0 \sqrt{ 1 + {{3 A^2 \omega^2}\over{4 g L}}} = 
\omega_0 \sqrt{ 1 + {{\omega^2}\over{\omega_c^2}} } = 
\omega_0 \sqrt{ 1 + R} .
\end{equation}
We have used $\omega_0 = \sqrt{3g/2L}$, which is the angular frequency
of the undriven pendulum. Driving the pendulum increases its frequency
(decreases its period). 

\subsection{\label{sec:INV} Drive angle 180 degrees}

If $\theta_d = 180^\circ = \pi$, 
we let $\delta = \phi - \pi$ and obtain the driven inverted
pendulum from Eq.~(\ref{eqn:ddotphi}):
\begin{equation}
\ddot{\delta} + \big({{9 A^2 \omega^2}\over{8 L^2}} \cos\delta - 
{{3 g}\over{2 L}}\big) \sin\delta = 0 .
\end{equation}
Small $\delta$ allows us to make the approximations $\cos\delta \simeq 1$
and
$\sin\delta \simeq \delta$:
\begin{equation}
\ddot{\delta} + {{3 g}\over{2 L}} \big({{3 A^2 \omega^2}\over{4 g L}} - 1
\big) \delta = \ddot{\delta} + \omega_p^2 \delta \simeq 0,
\end{equation}
which gives small oscillations with an angular frequency
\begin{equation}
\omega_p = \omega_0 \sqrt{ {{3 A^2 \omega^2}\over{4 g L}} - 1}
= \omega_0 \sqrt{ {{\omega^2}\over{\omega_c^2}} -1}
= \omega_0 \sqrt{ R -1} .
\end{equation}
We obtain stable small oscillations only if
$R > 1$ or equivalently $\omega > \omega_c$.
A stable driven inverted pendulum has a lower frequency (longer period) 
of slow motion than a free pendulum with the same geometry.

The traditional treatment of the inverted pendulum starting with
Eq.~(\ref{eqn:motion}) is reviewed in the Appendix. The solution of the 
linearized equation of motion for
$\theta_d=180^\circ$ gives the Mathieu functions.\cite{AS} Only for a
limited range of drive amplitude and angular frequency do we obtain stable
oscillations of the driven inverted pendulum. The range of stability is
usually displayed in terms of Mathieu parameters (see
Fig.~\ref{fig:aq}).

Figure~\ref{fig:Aw} displays the regions of stability of the Mathieu 
functions plotted in terms of the pendulum drive parameters
$A$ and $\omega$ with $L=0.25$\,m.\cite{m17q} The lower smooth curve in
Fig.~\ref{fig:Aw} is approximately 
$A\omega = 1.81$\,m/s. 
Note that we should avoid large driving amplitudes ($A
> 0.08$\,m) for a 25\,cm rod. The effective potential technique of
Ref.~\onlinecite{Landau} is not valid for large excursions from
equilibrium, so does not address an upper stability limit. A
numerical solution of Eq.~(\ref{eqn:motion}) is needed to explore
shorter rods (or large driving amplitudes) together with slow 
oscillation angular amplitudes larger than the linear
approximation supports.

\begin{figure}[h]
\includegraphics[width=0.60\textwidth]{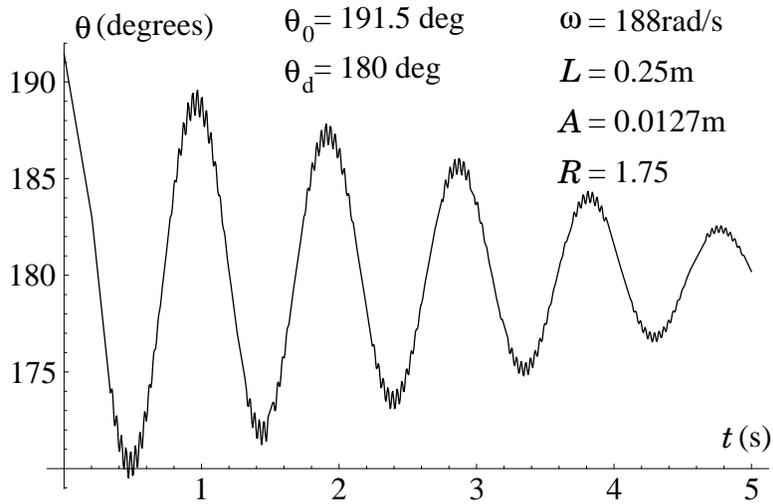}
\caption{\label{fig:samp} Motion for $\theta_d = 180^\circ$ with
constant frictional damping.}
\end{figure}

A full numerical solution of Eq.~(\ref{eqn:motion}), including
\textit{ad hoc} frictional damping, for $\theta_d = 180^\circ$ is
shown in Fig.~\ref{fig:samp}. Constant friction gives a rapid
linear decrease in the oscillation amplitude, while viscous
damping gives the familiar under-damped oscillation with an
exponentially decreasing amplitude. Our demonstration pendulums
seem to be better described by constant friction.

Figure~\ref{fig:bad} shows interesting, non-oscillatory behavior
which appears in numerical solutions for very short rods. 
This possibly chaotic non-oscillatory behavior of the driven pendulum
suggests other studies of
overdriven parametric systems which can be simply related to a
driven mechanical pendulum. Previous
studies\cite{BD, RLK} could easily be extended to arbitrary drive
angles using the analysis in Sec.~\ref{sec:genang}.

\subsection{Drive angle 90 degrees}

For a drive angle $\theta_d = 90^\circ = \pi/2$, the equation of 
motion for slow oscillations, Eq.~(\ref{eqn:ddotphi}), reduces to
a simple form:
\begin{equation}
\ddot{\phi} + \big({{3 g}\over{2 L}} - 
{{9 A^2 \omega^2}\over{8 L^2}} \cos\phi \big) \sin\phi = 0.
\label{eqn:90}
\end{equation}
For small angles $\phi$ ($\cos\phi \sim 1$ and $\sin\phi\sim
\phi$), we obtain
\begin{equation}
\label{temp1}
\ddot{\phi} + {{3 g}\over{2 L}}( 1 - R) \phi \simeq 0 .
\end{equation}
Equation~(\ref{temp1}) gives small oscillations near $\phi = 0$ only if
$\omega < \omega_c$ or $R<1$. However, if $\omega > \omega_c$
($R>1$), the term in parentheses in Eq.~(\ref{temp1}) is negative,
and there cannot be stable oscillation near $\phi=0$. To find the
new location of the stable equilibrium angle ($\phi_0$), we take
$\phi = \delta + \phi_0$ in Eq.~(\ref{eqn:90}) and write
\begin{equation}
\ddot{\delta} + {{3 g}\over{2 L}}
[ 1 - R \cos(\delta + \phi_0)]
\sin(\delta + \phi_0) = 0.
\label{eqn:nonl90}
\end{equation}
We obtain the equilibrium angle $\phi_0$ by taking $\delta = 0$ and 
$\ddot{\delta} = 0$:
\begin{equation}
(1 - R \cos\phi_0)
\sin\phi_0 = 0 .
\label{eqn:90equil}
\end{equation}
The solution $\phi_0 = 0$, which we have already seen, is 
stable only if $\omega < \omega_c$ ($R<1$).
Equation~(\ref{eqn:90equil}) also has a second solution valid for $R>1$:
\begin{equation}
1 - R \cos\phi_0 = 0 .
\end{equation}
The equilibrium angle $\phi_0$ is nonzero and finite for $R>1$:
\begin{equation}
\phi_0 = \cos^{-1} {1\over R}.
\label{eqn:90ang}
\end{equation}
Observe that $0 < \phi_0 \le \pi/2$ for $R>1$
(or $\omega > \omega_c$).

Near equilibrium this second solution should reduce
Eq.~(\ref{eqn:nonl90}) to an approximate harmonic oscillator:
\begin{equation}
\label{temp2}
\ddot{\delta} + \omega_p^2 \delta \simeq 0.
\end{equation}
Equation~(\ref{temp2}) gives a stable equilibrium only if the
coefficient of
$\delta$ is positive.
\begin{equation}
{d \over{d\delta}}\big\{ [ 1 - 
R \cos(\delta + \phi_0)]
\sin(\delta + \phi_0)\big\}_{\delta = 0} > 0,
\end{equation}
which yields
\begin{equation}
[1 - R \cos\phi_0 ] \cos\phi_0 + R \sin^2\phi_0 > 0.
\label{eqn:90stab}
\end{equation}
The requirement for equilibrium at $\phi_0$ in 
Eq.~(\ref{eqn:90equil}) makes the first term in
Eq.~(\ref{eqn:90stab}) zero. Hence
\begin{equation}
R \sin^2\phi_0 > 0,
\end{equation}
which holds for 
all $0 < \phi_0 \le \pi/2$ ($\omega > \omega_c$ or $R>1$).

\begin{figure}[h]
\includegraphics[width=0.40\textwidth]{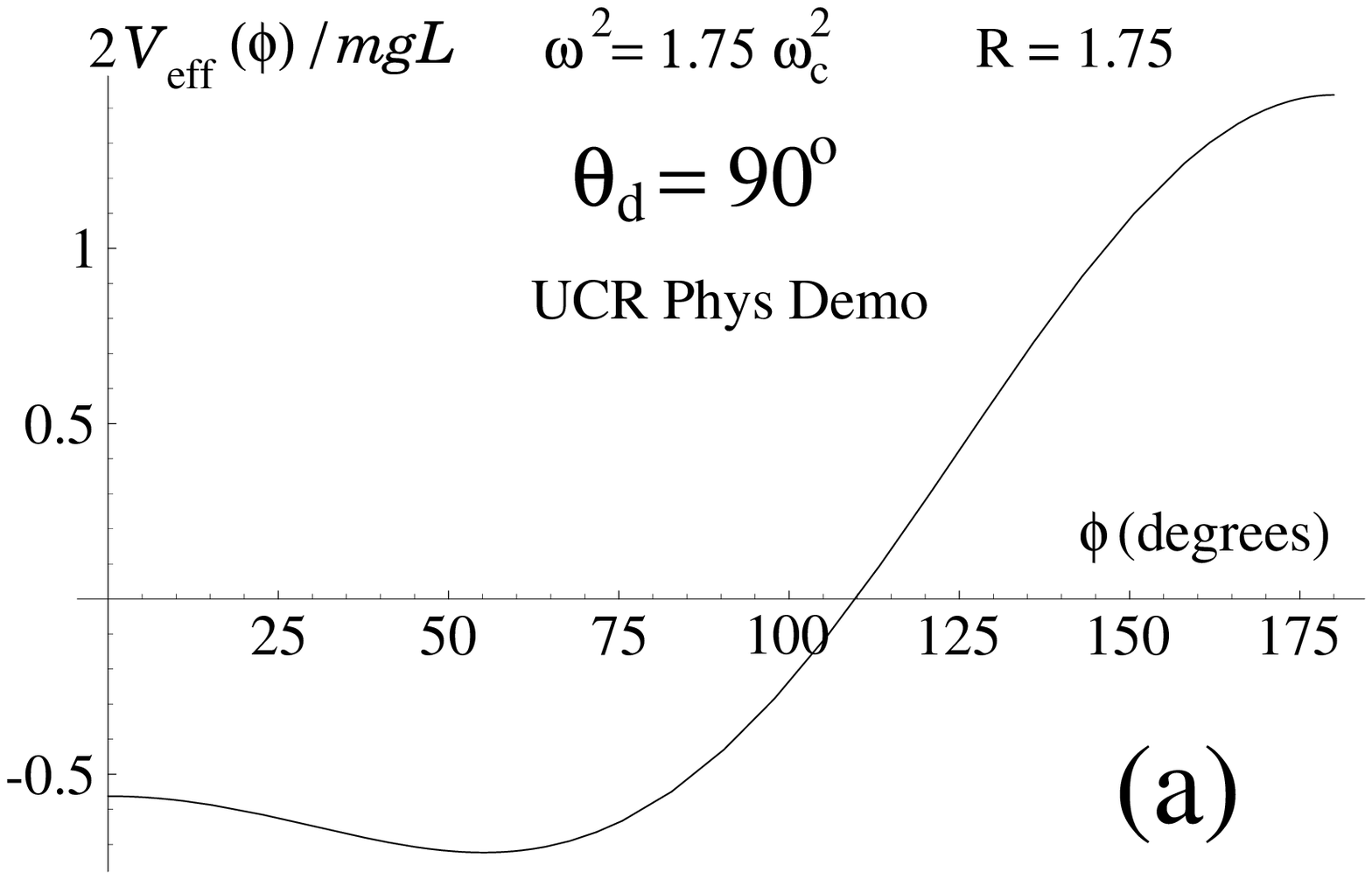} 
\hskip 0.1\textwidth
\includegraphics[width=0.40\textwidth]{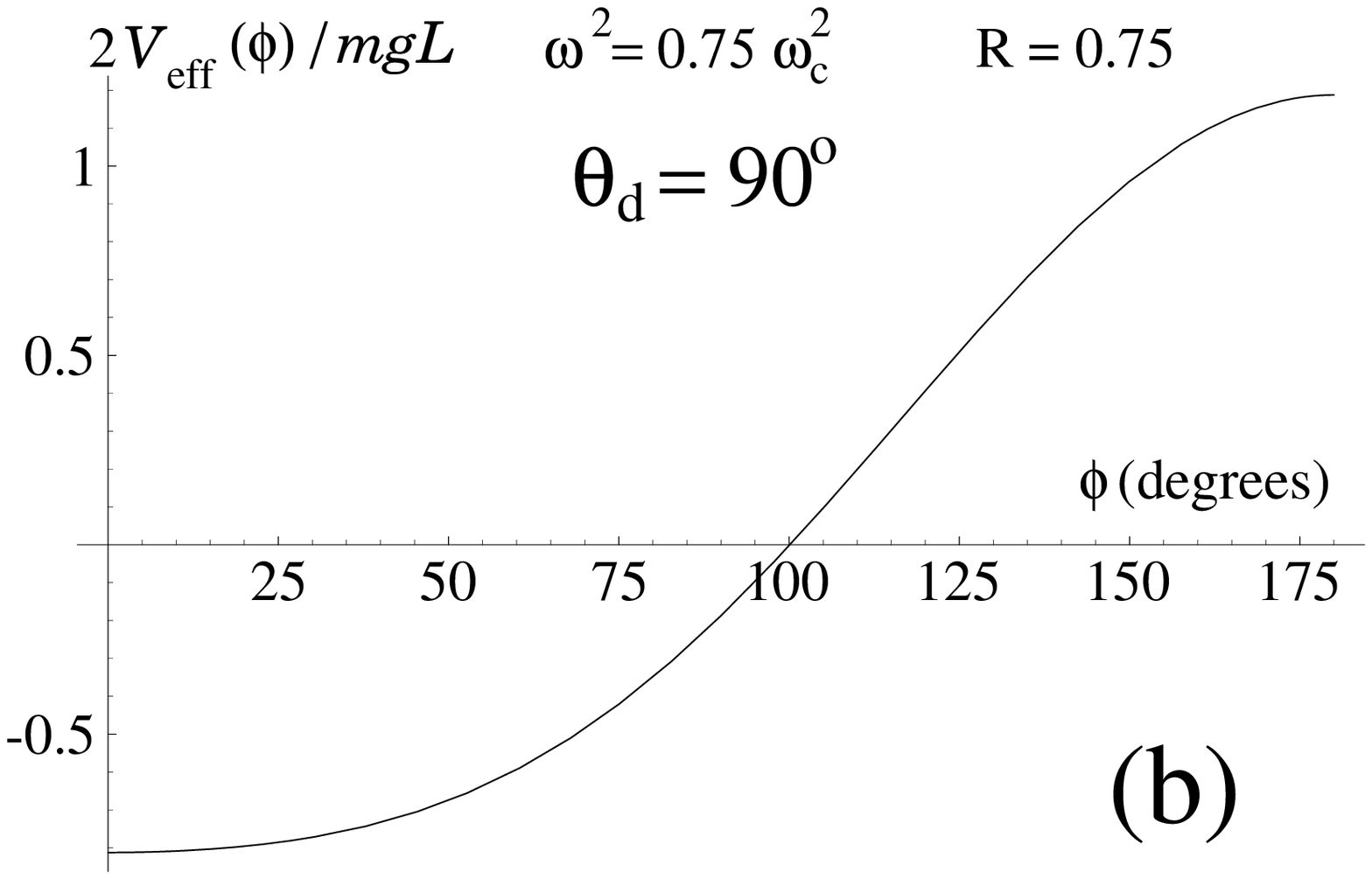}
\caption{\label{fig:90V} (a) Minimum of the effective potential
$V_{\text{eff}}(\phi)$ for $R = \omega^2/\omega_c^2 =
1.75$ at $\phi_0 \simeq 55^\circ$; the equilibrium at
$0^\circ$ becomes unstable. (b) For $R = 0.75$ ($\omega <
\omega_c$), the only minimum in
$V_{\text{eff}}(\phi)$ is at $\phi_0 = 0$.}
\end{figure}

Figure~\ref{fig:90V}(a) shows the effective potential for $R=1.75$
with a stable minimum at $\phi_0 = 55^\circ$. The effective
potential for $R=0.75 < 1$ is shown in Fig.~\ref{fig:90V}(b),
where the only minimum in the potential is at $\phi_0 = 0$.

The horizontally driven pendulum ($\theta_d = 90^\circ$) is stable
hanging down ($\phi_0 = 0$) for $R<1$, and stable near $\phi_0 =
\cos^{-1}(1/R)$ for $R>1$.

\section{\label{sec:genang} The Driven Pendulum at Any Angle}
We start with the equation of 
motion (\ref{eqn:ddotphi2}) for a general drive angle
$\theta_d$. 
\begin{equation}
\ddot{\phi} + {{3 g}\over{2 L}} \big( \sin\phi + 
{R\over 2} \sin[ 2 (\phi-\theta_d)]\big)
\simeq 0.
\end{equation}
Equilibrium occurs when $\phi = \phi_0$ with $\ddot{\phi} = 0$:
\begin{equation}
\sin\phi_0 + {R\over 2}
\sin [ 2(\phi_0 - \theta_d)] = 0.
\label{eqn:genang}
\end{equation}
As we have seen, the equilibrium is stable if 
\begin{equation}
{d \over d\phi }\big( \sin\phi + 
{R\over 2} \sin[ 2 (\phi-\theta_d)]
\big)_{\phi=\phi_0} > 0,
\end{equation}
which gives
\begin{equation}
\cos\phi_0 + R \cos [ 2(\phi_0-\theta_d)] > 0.
\label{eqn:anystab}
\end{equation}
Small oscillations about the equilibrium angle $\phi_0$ have the
angular frequency
$\omega_p$, where
\begin{equation}
\omega_p = \omega_0 \sqrt{ \cos\phi_0 + R
\cos [ 2(\phi_0-\theta_d)] } .
\label{eqn:omegap}
\end{equation}

The condition for equilibrium in Eq.~(\ref{eqn:genang}) can be easily
solved numerically for $\theta_d$ as a (multivalued) function of $\phi_0$.
However, we wish to know the stable angle of oscillation $\phi_0$ as a
function of the drive angle $\theta_d$. Simply finding the solutions of
Eq.~(\ref{eqn:genang}) is not sufficient. The zeroes must
correspond to real angles and to stable equilibriums as defined by
Eq.~(\ref{eqn:anystab}). Also, the ideal problem shown in
Fig.~\ref{fig:geom} has a two-fold ambiguity of drive angles at
$\theta_d$ and $\theta_d + 180^\circ$. Real lab or demonstration
apparatus will limit the motion of the rod so that we cannot
observe oscillations for
$|\theta_d - \phi_0| > 90^\circ$.

\begin{figure}[h]
\includegraphics[width=0.40\textwidth]{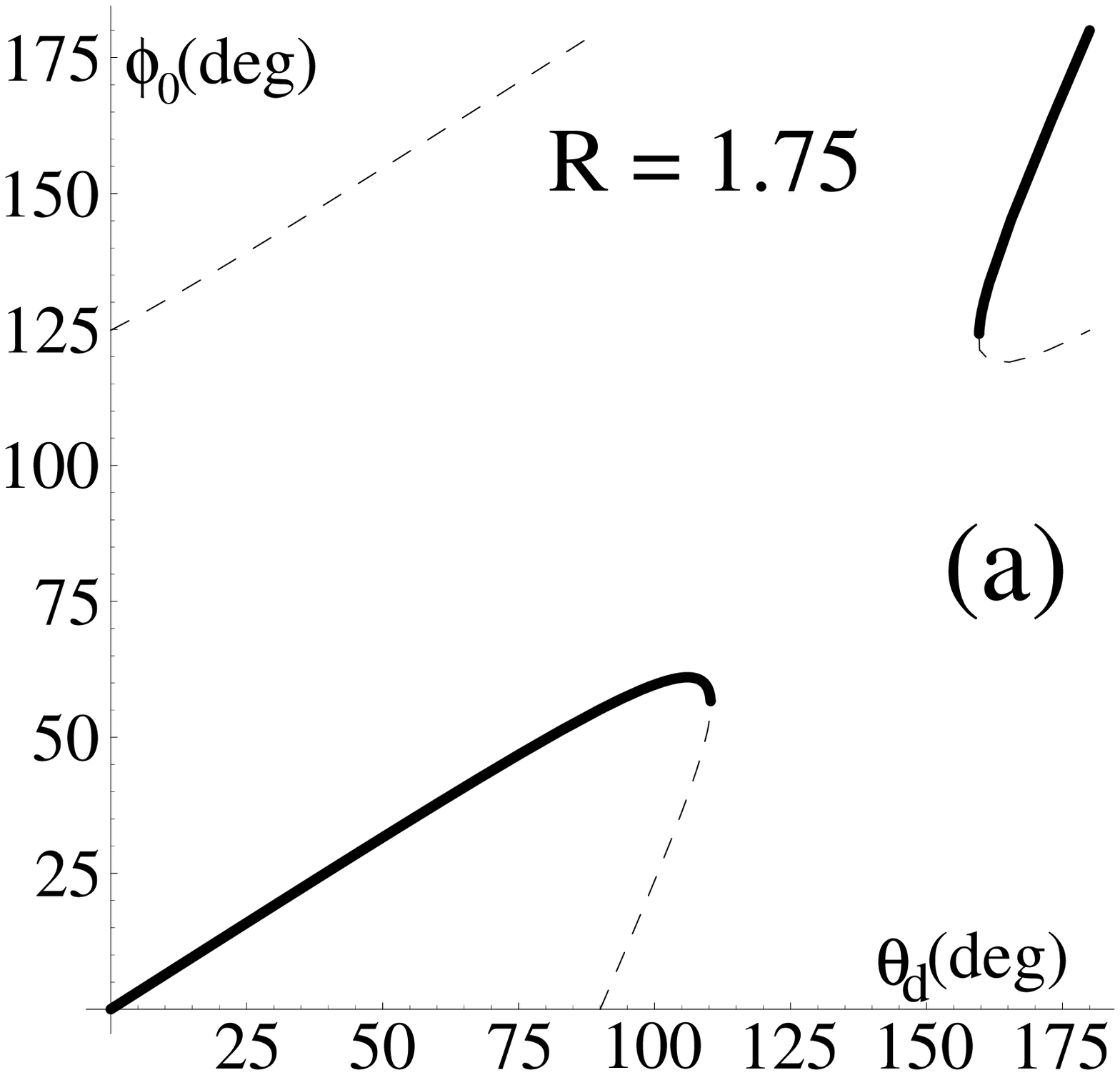} 
\hskip 0.1\textwidth
\includegraphics[width=0.40\textwidth]{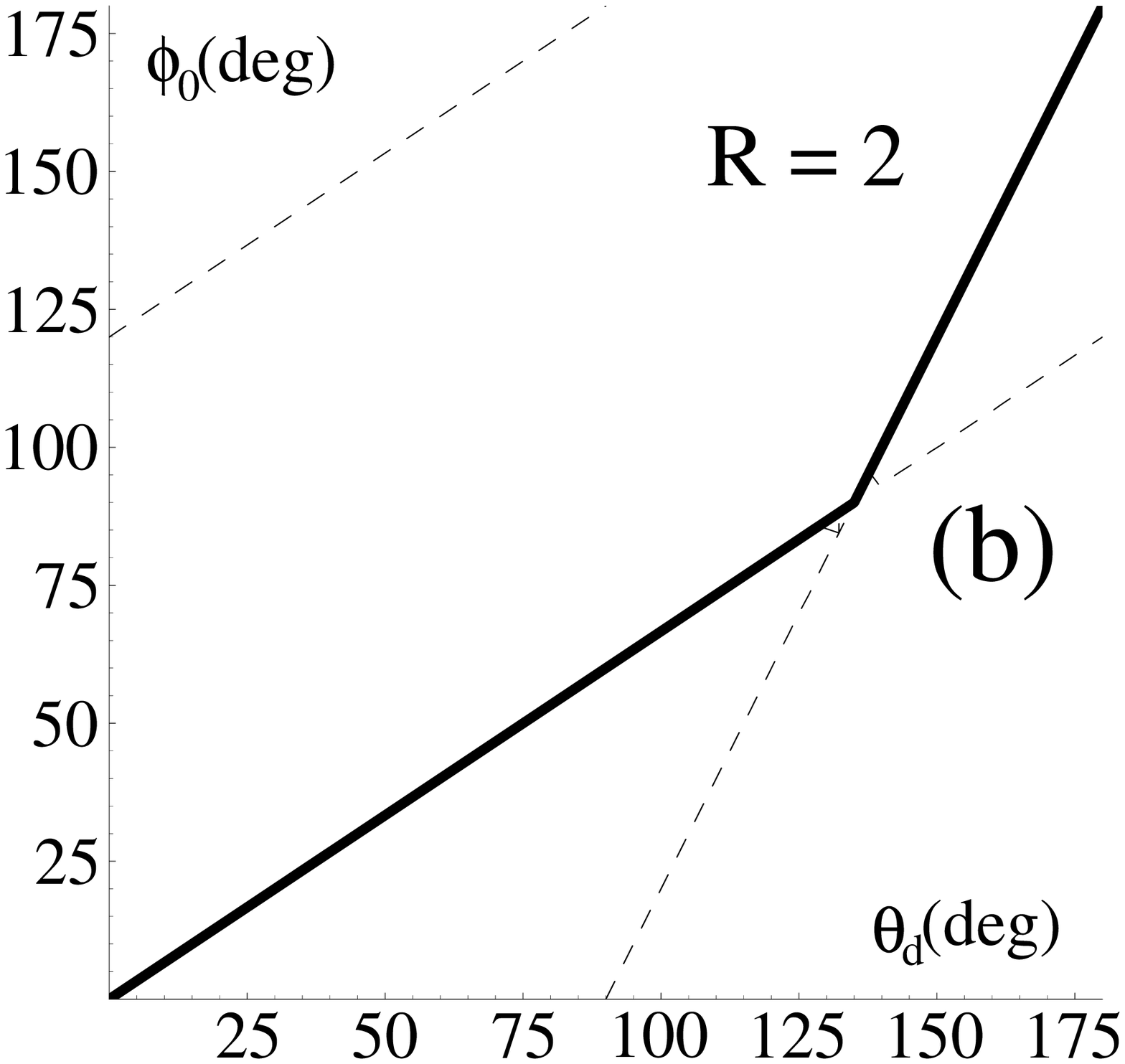}
\caption{\label{fig:equil} All equilibriums (dashed) and stable
equilibriums (solid line) are shown in $\phi_0$ versus $\theta_d$ for (a)
$R=1.75$, and (b) $R=2$.}
\end{figure}

Figure~\ref{fig:equil}(a) shows that a driven pendulum\cite{m17q}
with
$R=1.75$ has a range of drive angles for which there are no
(nearby) stable equilibrium angles. For all values of $R \ge 2$,
there is a stable equilibrium $\phi_0$ for all drive angles
$\theta_d$. The critical case
$R=2$ is shown in Fig.~\ref{fig:equil}(b). 

\begin{figure}[h]
\includegraphics[width=0.50\textwidth]{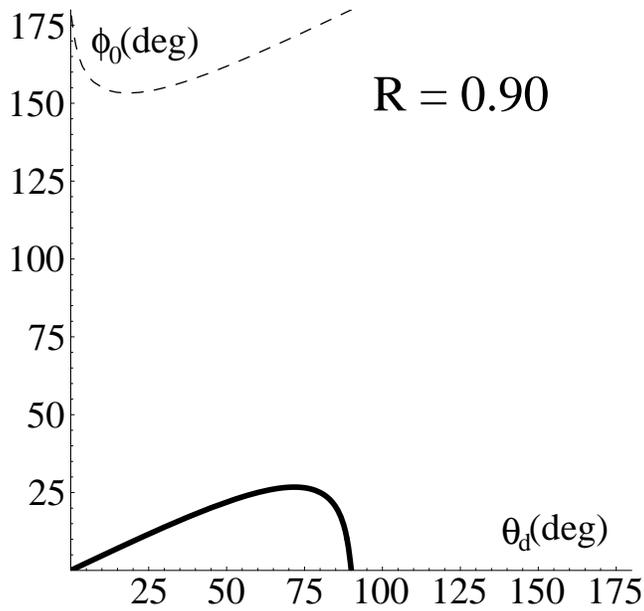}
\caption{\label{fig:nearone} All equilibriums (dashed) and stable
equilibriums (solid) are shown
in $\phi_0$ versus $\theta_d$ for $R=0.90$.}
\end{figure}

For values of $R>2$, the curve representing $\phi_0(\theta_d)$
becomes smooth, tending to a straight line as $R\to\infty$. (But we
must beware of an instability appearing at larger drive
parameters, equivalent to large
$R$. There are practical limits to how large $R$ can be to give a
stable driven pendulum in a physical system.) Values of $R$ less
than unity gives stable oscillations with 
$\theta_d \neq \phi_0$ over the driving angle range 
$0 < \theta_d < 90^\circ$ as shown in Fig.~\ref{fig:nearone}.

\section{Comparing Theory with Experiment for Any Drive Angle}
\begin{figure}[h]
\includegraphics[width=0.60\columnwidth]{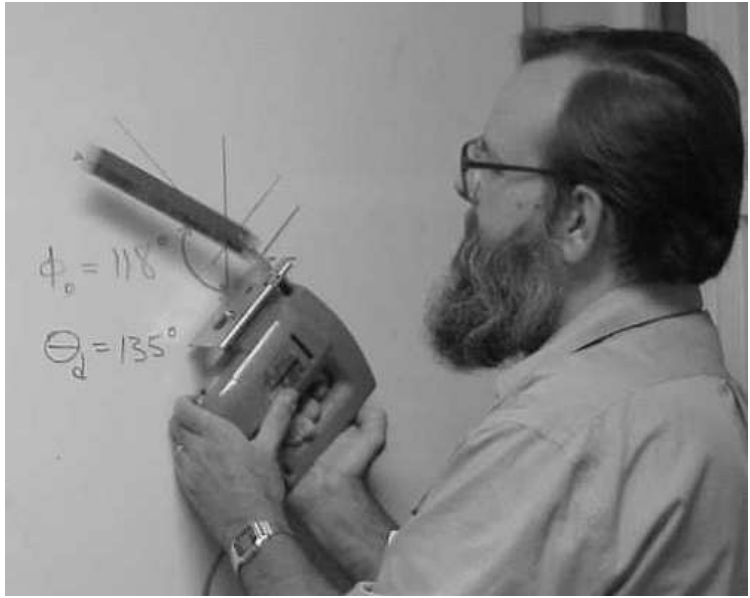}
\caption{\label{fig:ERAU2} A driven rod pendulum with $R=3.2$ and 
$\theta_d=135^\circ$ displaying stability at 
$\phi_0 \simeq 118^\circ$.}
\end{figure}

The desire for more degrees of freedom on a limited budget has led 
us to the driven pendulum demonstration using a small hand-held
variable speed jig saw (see 
Fig.~\ref{fig:ERAU2}). The conversion of the jig saw
parameters to the units used here gives the corresponding values
of $\omega$ and
$A$:
$0 < \omega < 330$\,rad/s and $A = 0.0089$\,m. The $L=20$\,cm
pendulum gives the expected maximum $R = 3.3$, well into the regime
where the driven pendulum is stable at all driving angles.

The value of $R$ can be measured using Eq.~(\ref{eqn:90ang}) for
$R > 1$. Hold the driving saw blade horizontally and measure the
angle to which the driven pendulum rises. At low driving speeds, it
will oscillate about the downward direction. Once $\omega >
\omega_c$, the pendulum will slowly rise with increasing drive
speed until a maximum angle is reached. Any angle above $60^\circ$
will correspond to
$R>2$, which then gives stable driven oscillations at any driving
angle. The 20\,cm pendulum has a measured maximum $\phi_0(90^\circ)
\simeq 72^\circ$, corresponding to $R \simeq 3.2$, in good
agreement with our expectation.

\begin{figure}[h]
\includegraphics[width=0.60\textwidth]{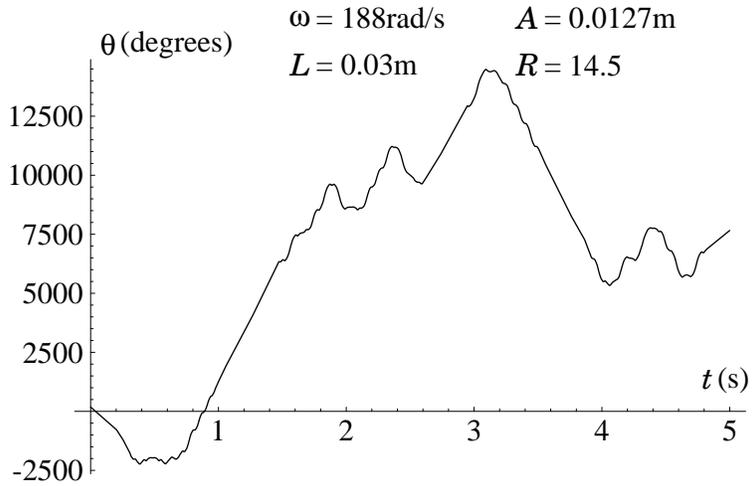}
\caption{\label{fig:bad} Over-driven inverted pendulum of the type
modeled in Fig.~\ref{fig:samp}, with $L=0.03$\,m. No
stable solution is found in the model or is observed
experimentally.}
\end{figure}

Figure~\ref{fig:ERAU2} shows the 20\,cm pendulum driven with
$\theta_d = 135^\circ$ and maximum frequency ($R\simeq 3.2$),
having being damped into a steady position near
$\phi_0(\theta_d = 135^\circ)
\simeq 118^\circ$. Shorter pendulum rods allows the exploration of
large $R$, where chaotic motion of the type shown in
Fig.~\ref{fig:bad} may be observed. From
simulations and experiments with shorter rods, we find that stable
driven oscillations do not extend to very large values of $R$. The
unstable system shown in Fig.~\ref{fig:bad} has a value of $R$ of
only 14.5. The upper limit on the amplitude from the Mathieu
function analysis described in the Appendix and shown in
Fig.~\ref{fig:Aw} does not reliably predict the onset of
instability, because the Mathieu equation only applies for small
drive amplitudes. The lower limit from the Appendix corresponds to
$R=1$. Establishing the correct upper limit on $R$ for stable
oscillation requires further study.

\section{Conclusions}
The pendulum driven at any angle from the vertical has been studied using
the effective potential method of Landau and Lifshitz\cite{Landau}
for rapid driving angular frequency ($\omega \gg \omega_p$).
Numerical simulations of the full equation of motion
(\ref{eqn:motion}) generally confirm the results of the simplified
model for 
$\omega \gg \omega_p$. The pendulum will oscillate about the
equilibrium angle
$\phi_0(\theta_d)$, defined by Eqs.~(\ref{eqn:genang}) and
(\ref{eqn:anystab}). The angular frequency of small oscillations
about equilibrium will occur at
$\omega_p$ given by Eq.~(\ref{eqn:omegap}).

The general behavior of the driven pendulum at any driving angle is
summarized by the parameter $R$ defined in Eq.~(\ref{eqn:Rdef}).
For $R \le 1$, there no stable inverted oscillations near
$180^\circ$. There are stable oscillations with $\phi_0>0$ for all
$0 < \theta_d < 90^\circ$ and
$\displaystyle \lim_{\theta_d \to 90^\circ} \phi_0 = 0$. For $1 <
R < 2$ there are stable inverted oscillations near $180^\circ$.
Some angles in the range $90^\circ< \theta_d <180^\circ$ are not
stable for $|\theta_d -
\phi_0|<90^\circ$. Also $\phi_0(90^\circ) > 0$. For $R \ge 2$,
there are stable oscillations for $|\phi_0 - \theta_d| <
90^\circ$ for all
$\theta_d$.

All parameters of the driven pendulum are accessible over interesting
ranges with simple and inexpensive apparatus. Even $g$ can be reduced by
tipping the plane of oscillation. We have explored more parameter sets than
reported here and hope that others will explore and report their own
variations of the driven pendulum at angles other than $0^\circ$ or
$180^\circ$.

\appendix
\section{Approximate Analytic Solution of the Inverted Pendulum}

Section~\ref{sec:INV} introduced the solution of the inverted pendulum based
on the effective potential approach of Ref.~\onlinecite{Landau}. This
system also can be easily solved by linearizing the equation of
motion as in Ref.~\onlinecite{DCM}.

For small angles the equation of motion for the driven inverted 
pendulum, Eq.~(\ref{eqn:motion}), simplifies to
\begin{equation}
\label{temp4}
\ddot{\delta} + \big( {{3 A\omega^2}\over{2 L}} \cos\omega t - 
{{3 g}\over{2 L}}\big)\delta \simeq 0.
\end{equation}
Equation~(\ref{temp4}) has the form of Mathieu's differential
equation\cite{AS}:
\begin{equation}
{{d^2y}\over{dz^2}} + ( a - 2q \cos 2z) y = 0.
\label{eqn:mathieu}
\end{equation}
If we make the substitution $z = \omega t/2$,
Eq.~(\ref{eqn:mathieu}) can be expressed in the Mathieu form,
\begin{equation}
{{d^2\delta}\over{dz^2}} + (- {{6 g}\over{L \omega^2}} + 
{{6 A}\over{L}} \cos 2z) \delta = 0 .
\label{eqn:matmot}
\end{equation}
The Mathieu parameters can be found by comparing
Eqs.~(\ref{eqn:mathieu}) and (\ref{eqn:matmot}) and are
$q = - 3 A/L$ and $a = - 6 g/(L\omega^2)$.

The general solutions of Mathieu's differential equation are
expressed by the even and odd Mathieu functions. For a portion of
the parameter space in
$(a,q)$, the Mathieu functions are real and periodic,
corresponding to stable equilibrium of the inverted driven
pendulum. Outside of this portion of the parameter space, the
functions are complex and divergent, corresponding to unstable
oscillations.

The region of parameter space that gives stable oscillations is 
defined through the Mathieu parameters $a$ and $q$.\cite{AS}
\begin{equation}
1 - |q| - {{q^2}\over 8} + {{|q|^3}\over{128}} - {{q^4}\over{1536}} - 
{{11 |q|^5}\over{36864}} + \cdots = a_1(q) > a .
\label{eqn:mup}
\end{equation}
\begin{equation}
a > a_0(q) = 
-{{q^2}\over 2} + {{7 q^4}\over{128}} - 
{{29 q^6}\over{2304}} + \cdots
\label{eqn:mlow}
\end{equation}
For the demonstration pendulum of Ref.~\onlinecite{m17q}, we have
$q = - 3 A/L = -0.1524$, so $q^2 \ll 1$. Then, to a good
approximation the lower limit from Eq.~(\ref{eqn:mlow}) simplifies
to the leading term.
\begin{equation}
a = -{{6 g}\over{L \omega^2}} > -{{q^2}\over 2} = -{{(3 A/L)^2}\over 2}.
\end{equation}
If we solve for the minimum angular frequency for stable inverted
driven oscillations, we obtain the familiar result of
Eq.~(\ref{eqn:Rdef}):
\begin{equation}
\omega > \omega_c = \sqrt{ {{4 g L} \over {3 A^2}}}.
\end{equation}
Including the $q^4$ correction from Eq.~(\ref{eqn:mlow}) increases
$\omega_{\rm c}$ by less than 1\%. The actual apparatus does not warrant
this level of precision, so we can safely devise a stable driven inverted
pendulum using $A\omega > \sqrt{ 4 g L/3}$.

\begin{figure}[h]
\includegraphics{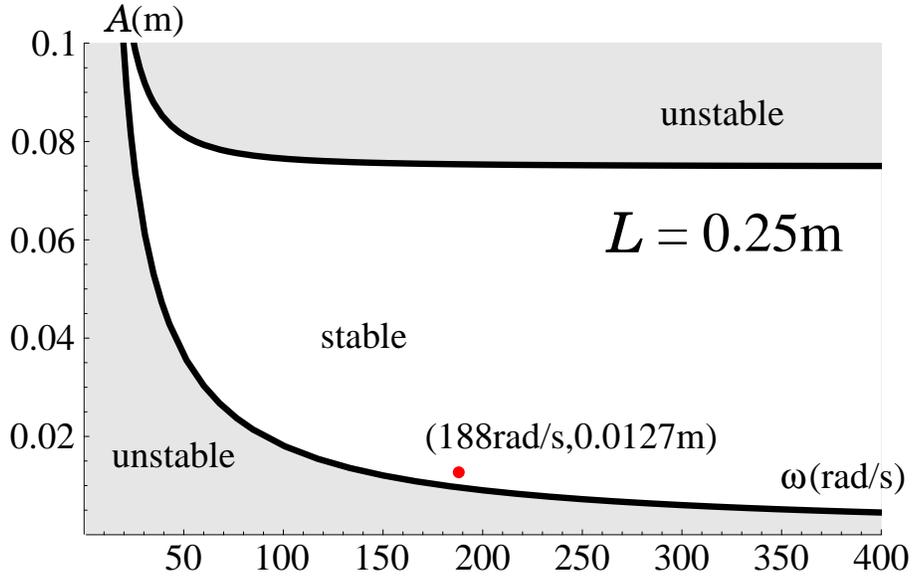}
\caption{\label{fig:Aw} The stable region in the drive amplitude
$A$ and angular frequency $\omega$ for a driven inverted pendulum
of length $L = 0.25$\,m. The parameters of the pendulum of
Ref.~\onlinecite{m17q} are shown as a point at $A=0.0127$\,m and
$\omega = 188$\,s$^{-1}$.}
\end{figure}

\begin{figure}[h]
\includegraphics{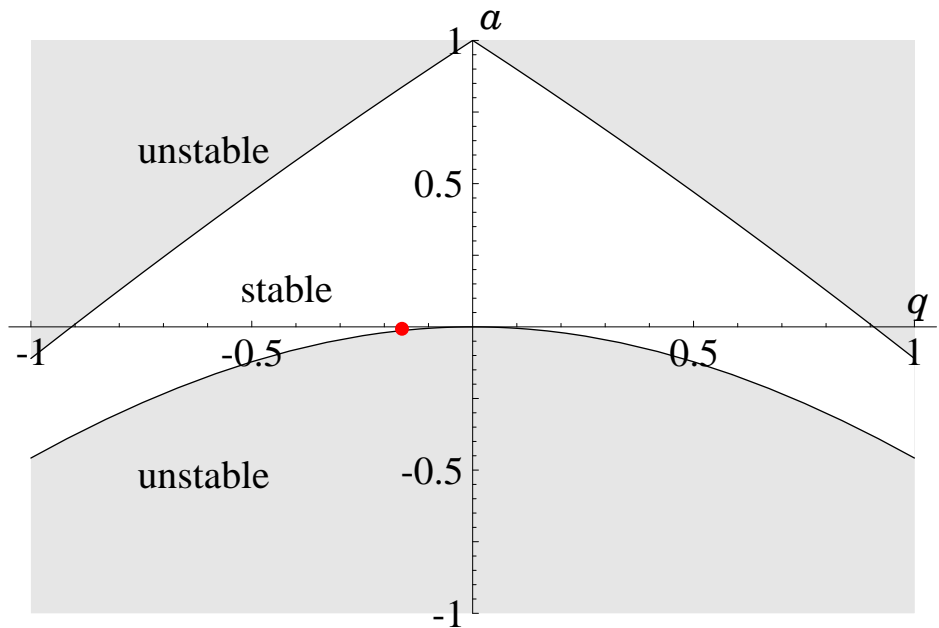}
\caption{\label{fig:aq} The stable region in the Mathieu parameters $a$ and
$q$ for a driven inverted pendulum.
The parameters of the pendulum of Ref.~\onlinecite{m17q} are shown as a
point at $q = -0.00665$ and $a = - 0.1524$.}
\end{figure}

Figure~\ref{fig:aq} is commonly used to display stable regions
in the Mathieu parameter space of $(a,q)$. It is more useful for our 
purposes to transform to the drive parameter space of
$(A,\omega,L)$. The
inversion of the first few terms of Eqs.~(\ref{eqn:mup}) and
(\ref{eqn:mlow}) for $a_0(q)$ and
$a_1(q)$ gives lower and upper limits for the drive amplitude $A$
as a function of the drive angular frequency $\omega$ and the
pendulum rod length
$L$.
\begin{equation}
A > A_0(\omega,L) = {L\over 3} 
\sqrt{ {{64 - \sqrt{ 4096 - 3584 {{6 g}\over{L
\omega^2}}}}\over{14}}}
\end{equation}
\begin{equation}
A < A_1(\omega,L) = {L\over 3}
\bigg( -2 + \sqrt{ 2} \sqrt{ 3 + {{6 g}\over{L \omega^2}}}\bigg).
\end{equation}
These equations were used to generate the limits of stability in the
parameter space $(A,\omega)$ for $L=0.25$\,m shown in
Fig.~\ref{fig:Aw}.

\begin{acknowledgments}
I would like to thank Ron Ebert of the University of California
Riverside for maintaining and documenting the many lecture
demonstrations used in countless classes, and for first
introducing me to the driven inverted pendulum. Thanks also to
Prof.~Darrel Smith and Embry-Riddle Aeronautical University for
allowing me to spend a sabbatical with them, and for giving me the
time to write this paper. I would also like to thank the referee
for comments which greatly improved the paper. Finally, thanks to
Prof.~Jos\'e Wudka for pointing out the Landau and Lifshitz
section on parametric oscillations, for sharing my amusement with
these little toys, and for encouraging me to write this all down. 
\end{acknowledgments}

\newpage


\end{document}